\UseRawInputEncoding
\documentclass[pdflatex,sn-mathphys]{sn-jnl}
\usepackage{siunitx}
\usepackage{xr}
\usepackage{lineno}
\usepackage{upgreek}
\usepackage{braket}
\usepackage{soul}
\usepackage{tabularx}
\usepackage{colortbl}
\DeclareSIUnit{\molar}{M}
\DeclareSIUnit{\Gauss}{G}
\DeclareSIUnit{\Tesla}{T}
\DeclareSIUnit{\Kelvin}{K}

\usepackage{booktabs}
\usepackage{adjustbox}
\usepackage{graphicx}
\usepackage{xcolor}
\renewcommand{\figurename}{Figure}

\begin{document}

\title[Extending Radiowave Frequency Detection Range]{Extending Radiowave Frequency Detection Range with Dressed States of Solid-State Spin Ensembles}


\author[1,2]{\fnm{Jens C.} \sur{Hermann}}
\equalcont{These authors contributed equally to this work.}
\author*[1,2]{\fnm{Roberto} \sur{Rizzato}}\email{roberto.rizzato@tum.de}
\equalcont{These authors contributed equally to this work.}
\author[1,3]{\fnm{Fleming} \sur{Bruckmaier}}
\author[1]{\fnm{Robin D.} \sur{Allert}}
\author[4]{\fnm{Aharon} \sur{Blank}}
\author*[1,2]{\fnm{Dominik B.} \sur{Bucher}}\email{dominik.bucher@tum.de}

\affil[1]{\orgdiv{Technical University of Munich, TUM School of Natural Sciences}, \orgname{Department of Chemistry}, \orgaddress{\street{Lichtenbergstra{\ss}e 4}, \city{Garching bei M{\"u}nchen}, \postcode{85748}, \country{Germany}}}
\affil[2]{\orgdiv{Munich Center for Quantum Science and Technology (MCQST)}, \orgaddress{\street{Schellingstr. 4}, \city{M{\"u}nchen}, \postcode{80799}, \country{Germany}}}
\affil[3]{\orgdiv{QuantumDiamonds GmbH}, \orgaddress{\street{Friedenstr. 6}, \city{M{\"u}nchen}, \postcode{81671}, \country{Germany}}}
\affil[4]{\orgdiv{Schulich Faculty of Chemistry, Technion  - Israel Institute of Technology}\orgaddress{\street{}, \city{Haifa}, \postcode{32000}, \country{Israel}}}

\abstract{Quantum sensors using solid-state spin defects excel in the detection of radiofrequency (RF) fields, serving various purposes in communication, ranging, and sensing. For this purpose, pulsed dynamical decoupling (PDD) protocols are typically applied, which enhance sensitivity to RF signals. However, these methods are limited to frequencies of a few megahertz, which poses a challenge for sensing higher frequencies. We introduce an alternative approach based on a continuous dynamical decoupling (CDD) scheme involving dressed states of nitrogen vacancy (NV) ensemble spins driven within a microwave resonator. We compare the CDD methods to established PDD protocols and demonstrate the detection of RF signals up to $\sim$ 85 MHz, about ten times the current limit imposed by the PDD approach under identical conditions. Implementing the CDD method in a heterodyne synchronized protocol combines the high frequency detection with high spectral resolution. This advancement extends to various domains requiring detection in the high frequency (HF) and very high frequency (VHF) ranges of the RF spectrum, including spin sensor-based magnetic resonance spectroscopy at high magnetic fields.}

\maketitle

%
\section*{Introduction}\label{sec1}
The detection of weak radiofrequency (RF) magnetic fields by spin-based magnetometers has been shown to be of importance for many applications, ranging from fundamental physics \cite{Sushkov_PhysRevA.108.010101,Budker_PhysRevLett.126.141802}, communications \cite{magaletti_quantum_2022,chen_quantum_2023}, and chemical analysis \cite{devience_nanoscale_2015,lovchinsky_nuclear_2016,glenn_high-resolution_2018,allert_advances_2022,rizzato_quantum_2023}. In particular, quantum sensors based on ensembles of spin defects in solid-state systems have shown increased sensitivity \cite{devience_nanoscale_2015,henshaw_nanoscale_2022, smits_two-dimensional_2019,bucher_hyperpolarization-enhanced_2020,liu_surface_2022, Allert_2022} and imaging capabilities \cite{bruckmaier2023imaging,briegel_optical_2024} in detecting weak signals, even from nuclear spins in micro- and nanoscale sample volumes. 
Typical sensing protocols rely on pulsed dynamical decoupling (PDD) methods that effectively decouple the sensor spins from the environmental noise, thereby increasing sensitivity to RF signals. 
Specifically, PDD methods use a train of microwave (MW) pulses, where the time intervals between pulses must be precisely matched to the RF frequency of interest \cite{Carr_Purcell_1954,meiboom_modified_1958,maudsley_modified_1986,Hanson_Science_doi:10.1126/science.1192739,abe_tutorial_2018,Bar-Gill,bar-gill_suppression_2012,pham_enhanced_2012,munuera-javaloy_dynamical_2021,degen_quantum_2017}. Although these sequences can significantly extend spin coherence times and provide high sensitivity for detecting low-frequency signals (sub-MHz to a few MHz), they ultimately fail when higher frequencies need to be detected \cite{degen_quantum_2017,Levine,wang_nanoscale_2021,wang_sensing_2022}. This is primarily constrained by the finite MW pulse width, determined ultimately by the available MW power \cite{PhysRevApplied.10.054059,Carlos_PhysRevLett.130.133603,daly2024nutationbasedlongitudinalsensingprotocols}. In addition, the use of a large number of MW pulses can lead to a significant accumulation of errors due to pulse imperfections \cite{Dobrovitski_comparison_2012,Tetienne,Bar-Gill,Souza,Plenio_PhysRevLett.122.200403}. Finally, from a technological perspective, the sampling rates of the waveform generators used to synthesize the microwave pulses can also be a practical limitation.
All of these factors represent significant limitations, especially in applications that require the detection of higher RF frequencies, such as in the case of spin sensor-based nuclear magnetic resonance (NMR). In this context, the use of stronger bias magnetic fields to improve intrinsic spectral resolution requires the detection of higher nuclear Larmor frequencies \cite{aslam_nanoscale_2017,Carlos_PhysRevLett.130.133603,Hu_10.1021/acs.nanolett.3c04822}.

Continuous dynamical decoupling (CDD) methods, such as spinlock (SL) - based sequences, detect RF fields by matching their MW amplitude rather than adjusting the MW pulse spacing, and offer a potential solution to these problems \cite{HartmannHahn,jeschke_coherent_1999,loretz_radio-frequency_2013,hirose_continuous_2012,scheuer_robust_2017,wang_nanoscale_2021}. Furthermore, the sensitivity of spinlock-based methods relies on the longitudinal relaxation time in the rotating frame ($T_{1\rho}$), which is typically much longer than the spin coherence time ($T_2$) that limits pulsed experiments \cite{wang2020coherence,wang_nanoscale_2021,rizzato_polarization_2022,rizzato_extending_2023}.
However, SL protocols necessitate precise control over the MW field amplitude and strong and spatially homogeneous microwave (MW) fields are particularly needed for NV-ensembles \cite{wang_sensing_2022,rizzato_polarization_2022}. Thus, it is crucial to achieve high MW field strengths to match the Rabi frequency with the RF signals and ensure high spatial homogeneity of the MW drive for effective spin-lock pulses over the spin defects ensemble. This makes the optimization of MW delivery critically important for the successful performance of CCD experiments.

In this work, we use ensembles of nitrogen vacancy (NV) centers in diamond and a resonant MW structure operating at $\sim$9.4 GHz (X-band) to demonstrate the detection of RF signals at frequencies up to $\sim$ 85 MHz, approximately an order of magnitude higher than the frequency detection limit imposed by PDD experiments under the same experimental conditions. 
Finally, we exploit the phase sensitivity inherent in the spin locking protocol to implement a novel CDD-based coherently averaged synchronized readout (CDD-CASR) detection scheme  \cite{wang_nanoscale_2021,schmitt_submillihertz_2017,boss_quantum_2017,glenn_high-resolution_2018}. This method allows the detection of RF signals with high spectral resolution and, with our modification, extends the detectable frequency range by an order of magnitude \cite{allert_advances_2022, rizzato_quantum_2023}.


\section*{Results}\label{sec2}
\subsection*{Theoretical Background}
\begin{figure*}[ht]
  \centering
  {\includegraphics[width=1\textwidth]{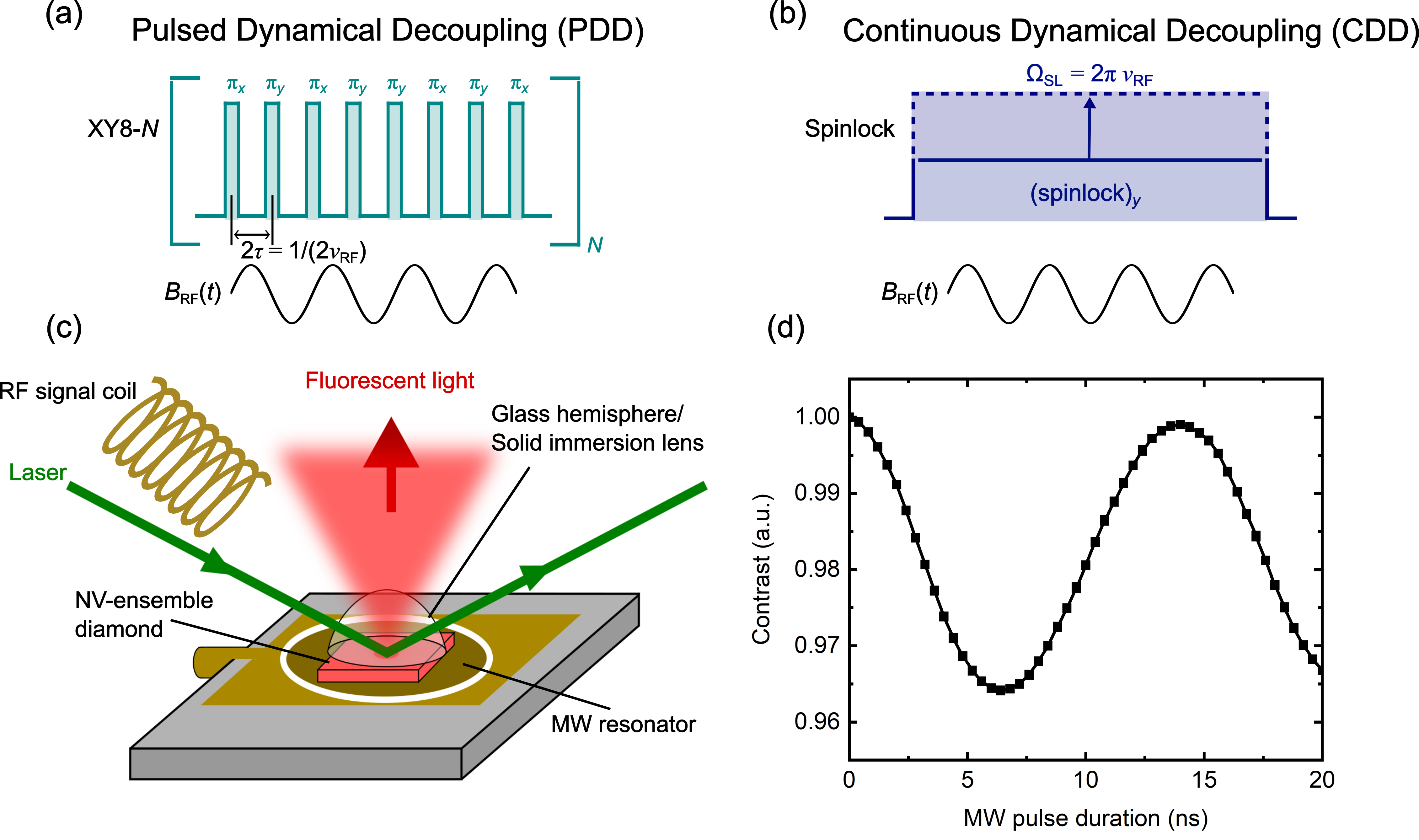}}
  \caption{\textbf{RF sensing with ensembles of NV-centers in a MW structure.} (a) In conventional Pulsed Dynamical Decoupling (PDD) sequences, such as XY8-$N$, the intervals between $\pi$ pulses must align with the period of the sensing RF. This requirement restricts their sensing capabilities, primarily due to pulse overlap and pulse errors. (b) In contrast, Continuous Dynamical Decoupling (CDD) sequences, such as the spinlock scheme, match the spinlock pulse amplitude to the sensing frequency, limiting their maximum detectable frequency to the highest achievable microwave amplitude, $\Omega_\mathrm{SL}$. (c) The experimental setup consists of a dielectric resonator with an NV diamond chip placed inside its cavity. This setup allows for high-power MW pulsing with a homogeneous magnetic field distribution over the NV ensemble area, making it an ideal platform for RF sensing using DD schemes. (d) With the assembly in (c), Rabi frequencies up to $\sim 85 \, \mathrm{MHz}$ can be achieved, corresponding to $\pi$-pulse durations of $\sim 6 \mathrm{ns}$.}
 \label{fig:Figure1}
\end{figure*}

In most quantum sensing schemes, the spin states of the NV centers are first initialized to the $\vert$0$\rangle$ state by a laser pulse and then transferred to a superposition state by a $\pi/2$ MW pulse. In this sensing state, alternating magnetic fields can be detected by selective manipulation of the spin state via MW pulses using pulsed (PDD) or continuous dynamical decoupling (CDD) schemes.
PDD methods, such as the XY8-$N$ sequence used in this work, consists of $N$ repetitions of a block of eight $\pi$ pulses around the $x$- and $y$-axis (see Figure \ref{fig:Figure1} (a)). In these conditions, a relative phase $\theta_\mathrm{PDD}(t_s)=(2/\pi)\gamma \hat{B}_\mathrm{RF} t_s$ between the $\vert$0$\rangle$ and $\vert$1$\rangle$ states is accumulated only when the matching condition $\nu_{\mathrm{RF}}=1/(4\tau)$ between the RF signal frequency $\nu_{\mathrm{RF}}$ and the spacing $2\tau$ between the $\pi$ pulses is met. Here, $\gamma$ is the gyromagnetic ratio of the electron, $\hat{B}_\mathrm{RF}$ the RF magnetic field amplitude and $t_s=16N\tau$ the interrogation time during which the phase $\theta$ is accumulated. To date, this has been the method of choice in NV-based quantum sensing, as it provides optimal sensitivities for signals up to a few MHz frequency and is particularly robust to pulse imperfections due to the alternate pulse phase design \cite{degen_quantum_2017,barry_sensitivity_2020,abe_tutorial_2018,smits_two-dimensional_2019,Bucher_principles,Liu_2022,bruckmaier2023imaging,briegel_optical_2024}.

Differently, the CDD sensing method relies on applying a long pulse along the y-axis (see Figure \ref{fig:Figure1} (b)) immediately after generating electron spin coherence with a strong $(\pi/2)_x$ pulse. This effectively locks the spins onto the equatorial plane of the Bloch sphere in the rotating frame (spinlock (SL)). In this protocol, the NV spins are brought into their dressed states (DS), where an artificial low-field condition is created and their pseudo-Zeeman splitting is determined by the Rabi frequencies of the bare states \cite{schweiger_principles_2001}. The lifetime of such a spin-locked state is limited by the spin-lattice relaxation in the rotating frame $T_{1\rho}$ which is typically longer than the decoherence times ($T_2$). This technique has been extensively used in NMR spectroscopy and in Dynamic Nuclear Polarization (DNP) experiments, where it enables polarization transfer between different nuclear spins, e.g. $^1$H and $^{13}$C \cite{HartmannHahn,levitt_spin_2013} and between spins at very different energy splittings, such as electron and nuclear spins \cite{henstra_nuclear_1988,weis_electron-nuclear_2006,Rizzato_MolPhys}. In addition, an arbitrary RF field with polarization transverse to the quantization axis of the dressed states can drive transitions between them and induce a rotating frame Rabi nutation whose dynamics encodes information about the sensing RF field \cite{jeschke_coherent_1999, loretz_radio-frequency_2013, wang_nanoscale_2021} (see Supplementary Note 1 in the Supplementary Information). In particular, during the SL time $t_s$, a phase factor $\theta_\mathrm{SL}=\frac{1}{2}\gamma \hat{B}_\mathrm{RF} t_s$ can be accumulated if the matching condition $\Omega_{\mathrm{SL}}=2\pi \nu_{\mathrm{RF}}$ is fulfilled. This pulse sequence has been successfully demonstrated with NV-centers in diamond and other spin defects in alternative materials \cite{loretz_radio-frequency_2013,degen_quantum_2017,wang_nanoscale_2021,rizzato_polarization_2022,rizzato_extending_2023,gong_coherent_2023}. 

\subsection*{Microwave delivery}
The experiment is based on a $1 \, \mathrm{mm} \times 1 \, \mathrm{mm}  \times 0.5 \, \mathrm{mm}$ electronic grade single crystal diamond chip on which an NV center-doped $10\, \upmu \mathrm{m}$ layer with ∼10 ppm substitutional nitrogen (P1 centers) was homoepitaxially grown (see “Methods” for further details). In order to achieve a strong and homogenous MW drive over the NV ensemble, we designed a dielectric resonator operating at a frequency of $9.4 \, \mathrm{GHz}$ (X-band), which is open on one side (see Figure \ref{fig:Figure1} (c), Methods and Supplementary Note 2 in the Supplementary Information) \cite{woflson_magnetic_2015}, in which the diamond can be inserted. This enables us to reach high Rabi frequencies of up to $\sim 85 \, \mathrm{MHz}$ and corresponding $\pi$-pulse durations of $\sim 6 \, \mathrm{ns}$ (see Figure \ref{fig:Figure1} (d) and Supplementary Note 3). In contrast, typical achievable Rabi frequencies are on the order of $20 \, \mathrm{MHz}$ with strong spatial inhomogeneities when the MW is delivered using simple microwave loops positioned on the NV-diamond surface \cite{bucherQuantumDiamondSpectrometer2019b,rizzato_polarization_2022, liu_surface_2022}. 
The assembly is positioned in the center of an electromagnet operating at a magnetic field strength $B_0$ of $0.3 \, \mathrm{T}$. This device combines precise MW pulsing, high-power MW generation, and uniform magnetic field distribution over the NV-ensemble. 
With this apparatus, we conduct RF sensing using a CDD protocol and compare its performance with a conventional PDD method based on an XY8-$N$ sequence. 

\subsection*{Testing the RF frequency response}
\begin{figure*}[ht]
  \centering
  {\includegraphics[width=0.9\textwidth]{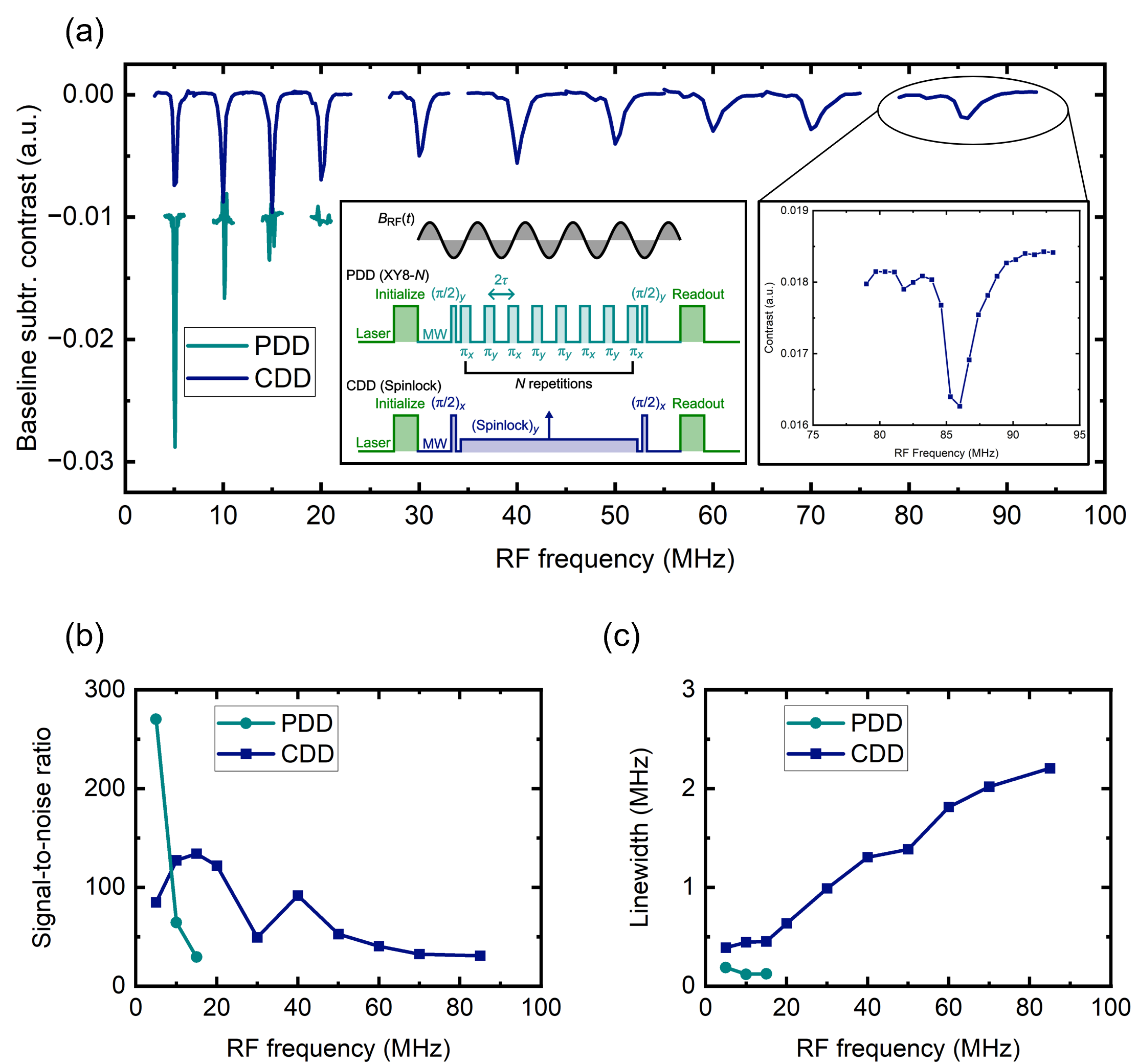}}
  \caption{\textbf{Extending the RF sensing range of NV ensemble based sensors.} 
  (a) Measured frequency responses of PDD (XY8-$N$) and CDD (spinlock) protocols (left inset) to an external RF field at different $\nu_\mathrm{RF}$. The pulse spacing $2\tau$ in case of PDD and the amplitude $\Omega_\mathrm{SL}$ of the spinlock pulse for the CDD are kept constant while the RF frequency is swept. The sensing limit ($\sim 10 \, \mathrm{MHz}$) of the PDD scheme can be overcome by CDD, ultimately achieving a maximum sensing frequency of $\sim 85 \mathrm{MHz}$ (right inset). 
  Signal-to-noise ratio (SNR) (b) and linewidths (c) of the corresponding data in (a).} 
 \label{fig:SL-vs-xy8}
\end{figure*}

In our first set of experiments, we test the maximum frequency $\nu_\mathrm{RF}$ of an RF field that the PDD and CDD protocols can detect. RF signals are generated by an antenna placed next to the diamond, and their frequency is measured by observing a dip in the NV's fluorescence contrast when the phase accumulation condition is met, as described above. We set a constant pulse spacing in PDD and a constant spinlock amplitude in the CDD case and keep the phase accumulation time the same for both experiments. By sweeping the applied RF frequency (see Figure \ref{fig:SL-vs-xy8} (a), inset), the frequency response of the NV sensor can be probed. 

 Figure \ref{fig:SL-vs-xy8} (a) shows the sensing of RF frequencies using an XY8-$N$ pulse sequence. The detected signal intensity decreases rapidly with increasing frequency $\nu_\mathrm{RF}$ and disappears ultimately around 20 MHz. We examined the pulse sequence in the time domain after the resonator, which reveals that the MW pulses are distorted and exhibit a tail of approximately 25 ns following the end of each pulse due to the MW ringing from the resonator (see Supplementary Note 4 in the Supplementary Information). This results in a pulse overlap for $\tau$-spacings on the order of 30 ns, consistently observed with sensing frequencies above 8 MHz. This demonstrates the difficulties with detecting high frequencies with PDD.
In contrast, using the CDD protocol, signals $\gtrsim 85 \, \mathrm{MHz}$ are succesfully detected. However, their signal-to-noise ratio (SNR) is three times lower than that of the XY8-$N$ experiments at low frequencies, when considering the peak intensity, with the noise remaining at a comparable magnitude for both pulse sequences. The integral under the peaks is comparable in both cases, indicating that the signal is distributed over a wider linewidth in the CDD case. This can be attributed to the microwave (MW) field inhomogeneity over the NV ensemble region, which is still suboptimal in our experimental setting. The trends in the linewidths are shown in Figure \ref{fig:SL-vs-xy8} (a) and (c). The recorded XY8-$N$ dips show narrower linewidths compared to the spinlock counterparts, although accurate linewidth estimation for signals recorded at frequencies $> 5 \, \mathrm{MHz}$ is challenging due to significant signal distortion. Nevertheless, it is apparent that, in the case of CDD, the linewidth gradually increases with higher RF frequencies (and MW amplitude), suggesting that possible heating effects may play a role. This might cause instabilities in the driving amplitude and thus lead to line broadening (see Supplementary Note 5 in the Supplementary Information for more details). Overall, while the XY8-$N$ protocol exhibits higher sensitivity compared to the spinlock, it is not suitable for detection above $10 \, \mathrm{MHz}$, whereas the CDD scheme is only limited by the achievable MW power which in our case corresponds to a maximum sensing frequency of $\sim 85 \mathrm{MHz}$. 


\subsection*{RF sensing beyond NV-decoherence time constraints}
\begin{figure*}[ht]
  \centering
  {\includegraphics[width=0.9\textwidth]{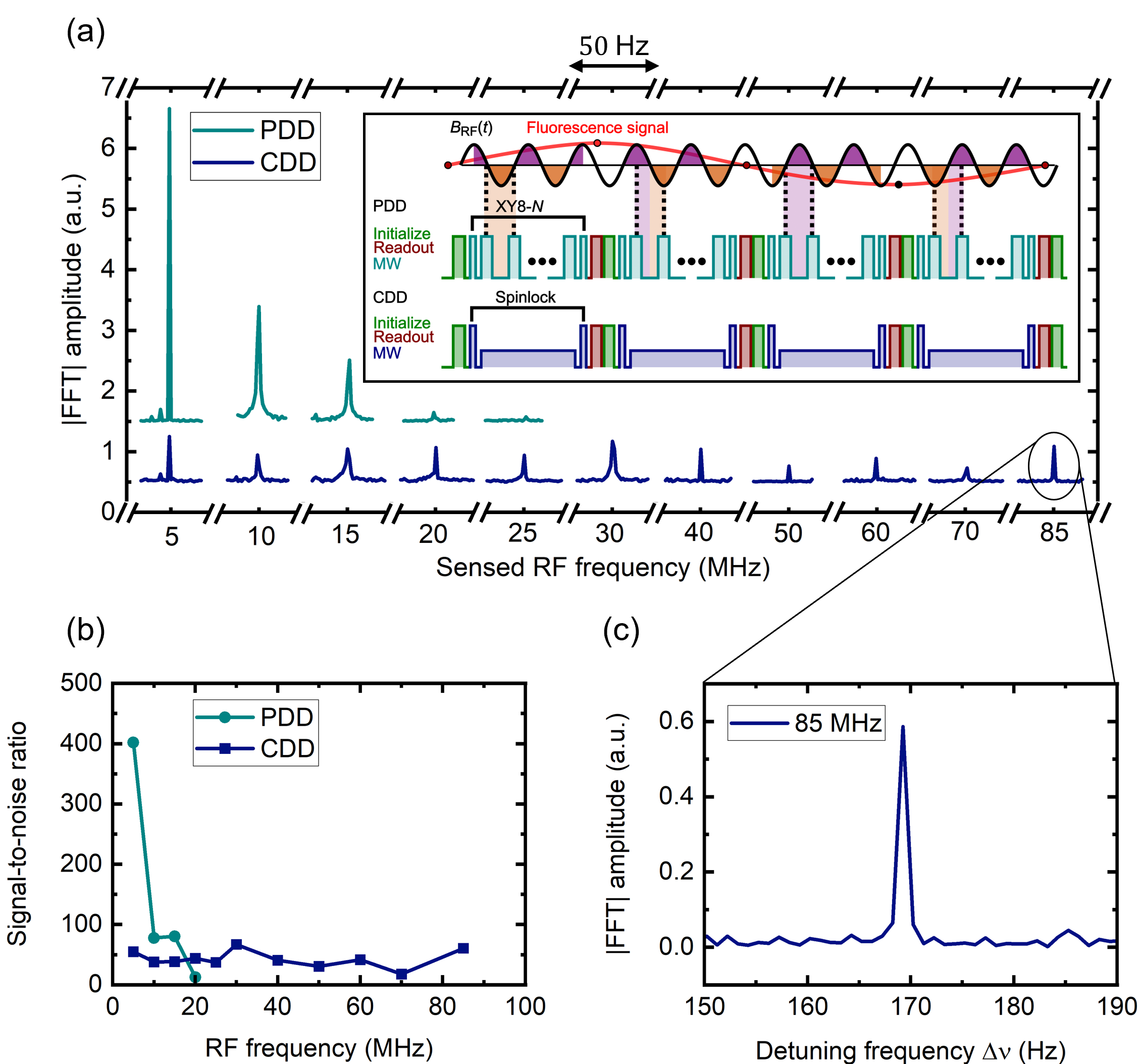}}
  \caption{\textbf{Overcoming limits on frequency resolution by Coherently Averaged Synchronized Readout (CASR).} (a) PDD (XY8-$N$) and CDD (spinlock) subsequences are implemented in the CASR scheme, as depicted in the inset. The time-domain fluorescence oscillations (schematically illustrated in the inset) are recorded for each RF sensing frequency up to 85 MHz (in case of CDD-CASR), and subsequently Fourier transformed, giving rise to the peaks at the detuning frequency $\Delta \nu$. Notice that the difference between two neighbouring $x$-axis breaks is $50 \, \mathrm{Hz}$ as indicated above the figure. (b) SNR of the CASR curves in (a). (c) Zoom-in of the CDD-CASR at an RF sensing frequency of $85 \, \mathrm{MHz}$ for a better visualization.}
  
 \label{fig:main-SR}
\end{figure*}

The aforementioned pulse sequences are limited in the achievable spectral resolution by the maximum sensing time which is set by their respective decoherence times ($T_2$ for XY8-$N$ and $T_{1\rho}$ for SL) or by technical contraints (such as MW inhomogeneity in CDD). To overcome this limitation, heterodyne \cite{schmitt_submillihertz_2017,boss_quantum_2017} or Coherently Averaged Synchronized Readout (CASR) \cite{glenn_high-resolution_2018} schemes have been developed, as illustrated in Figure \ref{fig:main-SR} (a). \\
In a typical CASR experiment, consecutive PDD sequences are set on resonance with the sensing RF frequency by adjusting the corresponding $\tau$ spacing according to the matching condition mentioned above. Then, the sequences are synchronized with the sensing RF signal and configured to be sensitive to the signal's phase (see Materials \cite{degen_quantum_2017}). This synchronization is achieved by setting the total duration of the individual PDD sequences to a multiple of the pulse spacing $\tau$. A slight deviation of the RF frequency $\nu_\mathrm{RF}$ with respect to the pulse sequence: $\nu_\mathrm{PDD}=1/(4\tau)$ results in different RF phases being recorded by each individual PDD sequence. As a consequence, the accumulated phases $\theta$, indicated by the violet and orange areas in the inset of Figure \ref{fig:main-SR} (a), differ for each subsequence. Thus, the fluorescence signal (red dots in Figure \ref{fig:main-SR} (a), inset) oscillates at the difference frequency $\Delta \nu=\vert\nu_\mathrm{PDD}-\nu_\mathrm{RF}\vert$ which can be Fourier transformed in order to obtain the corresponding frequency spectrum. Since the sensing time is not limited by the coherence time, the spectral resolution can be arbitrarily improved, only being limited by the clock stability of the experimental setup \cite{schmitt_submillihertz_2017,boss_quantum_2017,glenn_high-resolution_2018}.\\
Until now, this method has been implemented exclusively using PDD protocols, resulting in the same limitation regarding the maximum detectable frequency $\nu_\mathrm{RF}$. Our approach exploits the phase sensitivity of the spinlock scheme. By satisfying the condition $\nu_\mathrm{RF}=\nu_\mathrm{CDD}=\Omega_\mathrm{SL}/(2\pi)$ and introducing a small deviation between the pulse sequence and the RF signal synchronization, 
a novel CDD-CASR experiment with a wider frequency detection range is enabled. 
In the experiments shown in Figure \ref{fig:main-SR} (a), we compare the Fourier transformed signal obtained by PDD-CASR using the XY8-$N$ protocol with the CDD-CASR at different RF sensing frequencies. The detuning frequencies $\Delta \nu$  were between $150 - 200 \, \mathrm{Hz}$. A detailed summary of experimental parameters can be found in the Methods part. As in the previous result, at a sensing frequency of $\sim 5 \, \mathrm{MHz}$, the PDD-CASR exhibits a significantly higher SNR (see Figure \ref{fig:main-SR} b)) than the CDD-CASR. However, it rapidly declines for increasing RF frequencies whereas the peak heights of the CDD-CASR measurements remain fairly constant over all tested frequencies, enabling sensing of RF fields up to $\sim 85 \, \mathrm{MHz}$ (see Figure \ref{fig:main-SR} (c)) with high frequency resolution.
Although approximately eight times less sensitive than PDD at sensing frequencies as low as $\sim 5 \, \mathrm{MHz}$ (see Figure \ref{fig:main-SR} (b)), CDD-CASR demonstrates the ability to achieve sensitivities comparable to its counterpart at higher frequencies of around $\sim 10-15\, \mathrm{MHz}$. Importantly, it consistently performs well even at higher frequencies, proving to be a reliable method in a CASR protocol, with capabilities reaching up to $\sim$ 80-90 MHz.\\

\section*{Discussion}\label{sec3}
Our results highlight the CDD scheme as a complementary method to the well-established PDD for RF sensing, significantly extending the frequency detection range by an order of magnitude. The XY8-$N$ PDD sequence proves, in our case, effective for sensing low RF frequencies (a few MHz) until it becomes less effective above 5 MHz. Beyond this threshold, the spin-lock protocol demonstrates greater suitability due to its capability of sensing higher frequencies. However, the CDD scheme presents challenges that require further improvements in the future. Firstly, while it enables the detection of frequencies otherwise inaccessible, the method's sensitivity is around eight times lower than the established PDD approach when detecting frequencies as low as a few MHz (see Figure \ref{fig:main-SR} (b)). Since the theoretical sensitivity of the two methods is similar, the reduced CDD performance can be attributed to current experimental limitations, such as spatial inhomogeneities of the driving MW field and heating effects, both resulting in a broadening of the signal linewidth (see Figure \ref{fig:SL-vs-xy8} (c)). In particular, heating effects caused by high power and prolonged MW spinlock pulses, as detailed in Supplementary Note 5 of the Supplementary Information, likely represents the main limitation to overcome. It possibly contributes to two effects: 1) the instability of the microwave cavity resonance, which leads to a broader distribution of MW fields ($B_1$) on the experimental scale and subsequently widens the signal linewidth; 2) the increase in the diamond temperature itself, causing a shift in the NV resonance and degrading the performance of the sensing protocol. These issues can be effectively addressed through strategies for active cooling and a more careful design of the setup, incorporating a diamond sensor with a geometry precisely tailored to the resonant cavity to minimize the contribution from the MW $E$-fields. Furthermore, solving the heating problems would allow us to take advantage of the $T_{1\rho}$ time, which is typically much longer than the $T_2$ time limiting the PDD counterpart, resulting in improved sensitivity \cite{barry_sensitivity_2020,rizzato_polarization_2022,rizzato_extending_2023}.
Finally, methods for increased robustness of manipulating NV-dressed states need to be explored in this direction, including techniques based on continuous concatenated dynamical decoupling (CCDD) and double-drive experiments \cite{cai2012robust,stark_narrow-bandwidth_2017,cohen2017continuous,farfurnik2017experimental,wang2021observation,ramsay_coherence_2023,Luxmoore2024continuousdriveheterodynemicrowave}. Reducing the size of the MW structure for smaller spin ensembles down to the single spin level would alleviate the heating problems and reduce the technical complexity of implementation. 

In summary, the integration of Continuous Dynamical Decoupling (CDD) detection protocols with microwave (MW) resonators for spin ensembles extends the frequency detection range by an order of magnitude. While some technical challenges remain, in particular the optimal dissipation of heat generated by the high-power MW drive, these are primarily engineering tasks that are likely to be solved in the near future with more refined instrumental strategies. This improvement provides a valuable toolkit for quantum sensing based NMR, especially in high magnetic bias fields.

\clearpage
\section*{Methods}\label{sec4}
\textbf{NV ensemble sensor}\\
For the experiments, a $1\, \mathrm{mm} \times 1\, \mathrm{mm} \times 0.5\, \mathrm{mm}$ electronic grade diamond, provided by Element Six Ltd (Didcot, United Kingdom), was used. A nitrogen-doped layer was overgrown using chemical vapour deposition (CVD) by Applied Diamond, Inc. (Wilmington, DE, United States). The N$_{\mathrm{s}}^0$ concentration after the growth is estimated to be $\sim$ 10 ppm and the layer thickness is approximately $10 \, \upmu \mathrm{m}$. The diamond was irradiated with electrons with an energy of $1 \, \mathrm{MeV}$ and a fluence of $3 \cdot 10^{18} \, \mathrm{cm^{-2}}$ to create vacancies in the lattice. Subsequently, the diamond was annealed in vacuum for 12 hours at a temperature of 800$^{\circ}$C to form the NV centers.\\
\\
\textbf{Experimental setup}\\
As depicted in Figure \ref{fig:Figure1} (c), the diamond is placed into a dielectric microwave resonator operating in the X-band regime at $9.4 \, \mathrm{GHz}$. For an efficient light collection an optical glass hemisphere (TECHSPEC N-BK7, Edmund Optics) is glued on top of the diamond. After positioning this assembly inside an electromagnet (model 5405, GMW Associates) which is oriented such that the magnetic field $B_0$ aligns with one of the four NV axes, the NV center's $\vert$0$\rangle$ $\leftrightarrow$ $\vert$+1$\rangle$ transition frequency is tuned to the resonator eigenfrequency by adjusting the current through the electromagnet to a corresponding field of $\sim 230 \, \mathrm{mT}$. For initialization and readout of the NV qubit state, a laser (Opus 532, Novanta photonics) with a power of approximately $200 \, \mathrm{mW}$ is used.
Pulsing of the laser is enabled by an acousto-optic modulator (3250-220, Gooch and Housego) with a pulse duration of $1.1 \, \mathrm{ms}$. This long pulse duration is chosen to maximize the dead time between CDD sequences in order to mitigate possible resonator heating effects (see Supplementary Note 5 in the Supplementary Information). After passing a $\lambda/2$ waveplate which sets the polarization for optimal NV excitation, the laser light is focused  by a lens with a focal length of $75 \, \mathrm{mm}$ down to an estimated spot diameter of $\sim 50 \, \upmu \mathrm{m}$ on the NV diamond under a total reflection geometry. The fluorescence light is collected and collimated by a condenser lens (ACL25416U-B, Thorlabs) placed above the diamond and subsequently led through a long-pass filter (Edge Basic 647 Long Wave Pass, Semrock) and focused on a photodiode (PDA100A2, Thorlabs) by another condenser lens of the same model. The photodiode is then read out by a data acquisition unit (USB-6281 DAQ, National Instruments) which is connected to a computer.\\
An arbitrary waveform generator (AWG 5202, Tektronix) synchronizes the experiment and generates the microwave pulse sequences. An additional MW source (SMB 100A, Rohde\&Schwarz) acts as a local oscillator whose signal ($8.9 \, \mathrm{GHz}$) is mixed with that of the AWG ($0.5 \, \mathrm{GHz}$) via an IQ-mixer (MMIQ-0218LXPC, Marki Microwave) to achieve the desired $9.4 \, \mathrm{GHz}$ output frequency. After pre-amplification (ZX60-153LN-S+, Mini-Circuits), the MW signal reaches the $\sim 60 \, \mathrm{W}$ main amplifier (RFLUPA08G11GA, RF-Lambda) and thereafter the MW resonator.\\
The RF fields to be sensed in the experiments are delivered by placing a self-assembled copper coil near the diamond. The coil is soldered on a BNC cable and connected to an amplifier (LZY-22+, Mini-Circuits) and RF source (DG4162, RIGOL). To lock the RF signal to the pulse sequence in the CASR experiments, the RF source is synchronized with the AWG. In order for the coil to produce approximately the same magnetic field output for each sensing frequency, the RF amplitude was calibrated before the experiments, as described in Supplementary Note 6 in the Supplementary Information.\\
\\
\textbf{X-band microwave resonator}\\
The resonator was constructed from high permittivity ceramics ($\epsilon$= 80) and was machined into the shape of a ring with an outer diameter of 4.8 mm, an inner diameter of 2 mm, and a height of 1.55 mm. It is excited by a microstrip line positioned beneath it. The desirable resonance mode for our application is TE$_{01\delta}$ (see Figure 2 in Supplementary Note 2 of the Supplementary Information), which produces a high and homogeneous $\hat{B}_1$ in the center of the structure. For this mode, the resonance frequency is approximately 9.4 GHz, and the loaded quality factor $Q$ of the resonator, when critically coupled, is found to be $\sim 700$.\\
\\
\textbf{XY8-$N$ and spinlock protocols}\\
The XY8-$N$ experiments are conducted in accordance with the following scheme, $(\pi/2)_y \, [[-\tau - (\pi)_\phi - \tau]_8 ]_N \, (\pi/2)_y$, where the (relative) phases of the $\pi$-pulses are chosen such that the rotational axes alternate between the $x$- and $y$-axis: $\phi=[x$-$y$-$x$-$y$-$y$-$x$-$y$-$x]$ (see inset of Figure \ref{fig:SL-vs-xy8}). The durations of the $\pi/2$ and $\pi$ pulses are $3 \, \mathrm{ns}$ and $6 \, \mathrm{ns}$, respectively (see Supplementary Note 3 in the Supplementary Information). The first $\pi/2$ pulse and the last projection pulse are in phase ($y$-axis as rotational axis), enabling RF-phase insensitive variance detection (quadratic detection) which is suitable for sensing incoherent signals (see Supplementary Information of \cite{rizzato_extending_2023} for more details). For noise cancelling purposes, a second referencing sequence is implemented, only differing in the relative phase of the last projection pulse: $(\pi/2)_{(-y)}$. Each data point is then obtained by normalizing the difference between measurement and reference readouts to their sum and by averaging them 2,000 times.\\
In this experiment, the RF source generated a non-synchronized signal by keeping it unlocked in regards to the pulse sequence so that each repetition of the sequence results in a different RF phase being recorded. For each sensing frequency, the corresponding spacing between $\pi$ pulses is kept constant while the RF frequency is swept and the fluorescence signal is recorded. The phase accumulation time $t_s$, that is the timeframe between initial and last $\pi/2$ pulse is $5 \, \upmu \mathrm{s}$ for all experiments. This, together with the dead time of $1.1 \, \mathrm{ms}$ set by the laser pulse, results in a duty cycle of $\sim 0.5\%$. Such a low duty cycle is chosen to reduce heating effects of the resonator, which could negatively affect the performance of the spinlock experiment (see Supplementary Note 5 in the Supplementary Information).
\\
The spinlock experiment in variance detection mode consists of the following pulsing scheme: $(\pi/2)_x \, (\text{Spinlock})_y \, (\pi/2)_x$, as illustrated in the inset of Figure \ref{fig:SL-vs-xy8} (a). Referencing is achieved by changing the relative phase of the last projection pulse to $(\pi/2)_{(-x)}$.
The experiments are conducted in the same manner as the XY8-$N$ measurements with identical phase accumulation times and number of averages. For every sensing frequency, the amplitude $\Omega_\mathrm{SL}$ is set accordingly, and the RF frequency is swept. \\
\\
\textbf{CASR protocol}\\
The CASR protocol consists of concatenated DD-subsequences, whereby the total duration of each subsequence is set to a multiple of the pulse spacing $\tau$. The RF frequency to be sensed slightly deviates from the matching condition, such that the obtained contrast oscillates with the detuning frequency $\Delta \nu=\vert\nu_\mathrm{DD}-\nu_\mathrm{RF}\vert$. In this experiment, detuning frequencies between $150-200 \, \mathrm{Hz}$ are chosen. In case of the PDD sequence, the spacing $\tau$ is set to a desired matching frequency $\nu_\mathrm{PDD}=1/(4\tau)$ (e.g. $5 \, \mathrm{MHz}$) and the RF signal is adjusted according to $\nu_\mathrm{RF}=\nu_\mathrm{PDD}+\Delta \nu$. For the CDD protocol, a sweep of the spinlock amplitude $\Omega_\mathrm{SL}$ is carried out while keeping the RF frequency constant at a chosen matching frequency. After identifying and setting the SL amplitude to the desired measurement frequency $\nu_\mathrm{RF}=\nu_\mathrm{CDD}=\Omega_\mathrm{SL}/(2\pi)$ at the fluorescence dip, a small deviation between the pulse sequence and RF synchronization is introduced, leading to the oscillating contrast in the time-domain.
\\
The CASR protocol requires sensitivity to the phase of the coherent RF signal, hence the last $\pi/2$ projection pulse of the subsequences differ from the previous experiment: $(\pi/2)_x$ and $(\pi/2)_y$ for the XY8-$N$ and spinlock subsequences, respectively. This sets the scheme to a slope detection mode (linear detection) which is sensitive to coherent signals (see Supplementary Information of \cite{rizzato_extending_2023} for more details). The phase accumulation time of the subsequences is kept at $5 \, \upmu \mathrm{s}$ and the total duration of the measurement is $1 \, \mathrm{s}$. The time signal is Fourier transformed and the resulting frequency spectrum $\mid \mathrm{FFT} \mid$ is plotted in Figure \ref{fig:main-SR}.


\bibliography{bibliography}

\section*{Acknowledgment} 
This study was supported by the European Research Council (ERC) under the European Union's Horizon 2020 research and innovation program (grant agreement No 948049), the European Union’s Horizon Europe research and innovation program under grant agreement No 101135742 and by the Deutsche Forschungsgemeinschaft (DFG, German Research Foundation) – 412351169 within the Emmy Noether program. D. B. B. acknowledges support by the DFG under Germany's Excellence Strategy – EXC 2089/1 – 390776260 and the EXC-2111 390814868. R.R. acknowledges support from the DFG Walter Benjamin Programme (Project RI 3319/1-1).


\section*{Author Contributions}

R.R. and D.B.B. designed the research. J.C.H built the experimental setup and carried out the experiments. F.B. programmed the pulse sequences. R.D.A. prepared the diamond sensor. A.B. designed and fabricated the MW resonator. R.R., J.C.H and D.B.B. analyzed the data. R.R., J.C.H. and D.B.B. wrote and reviewed the manuscript with inputs from all authors. 

All correspondence and request for materials should be addressed to R.R.(roberto.rizzato@tum.de) or D.B.B. (dominik.bucher@tum.de).


\section*{Competing interests} All authors declare that they have no competing interests.


\end{document}


\title[Supplementary Information]{Supplementary Information to ``Extending Radiowave Frequency Detection
Range with Dressed States of Solid-State Spin
Ensembles''}


\author[1,2]{\fnm{Jens C.} \sur{Hermann}}
\equalcont{These authors contributed equally to this work.}
\author*[1,2]{\fnm{Roberto} \sur{Rizzato}}\email{roberto.rizzato@tum.de}
\equalcont{These authors contributed equally to this work.}
\author[1,3]{\fnm{Fleming} \sur{Bruckmaier}}
\author[1]{\fnm{Robin D.} \sur{Allert}}
\author[4]{\fnm{Aharon} \sur{Blank}}
\author*[1,2]{\fnm{Dominik B.} \sur{Bucher}}\email{dominik.bucher@tum.de}

\affil[1]{\orgdiv{Technical University of Munich, TUM School of Natural Sciences}, \orgname{Department of Chemistry}, \orgaddress{\street{Lichtenbergstra{\ss}e 4}, \city{Garching bei M{\"u}nchen}, \postcode{85748}, \country{Germany}}}
\affil[2]{\orgdiv{Munich Center for Quantum Science and Technology (MCQST)}, \orgaddress{\street{Schellingstr. 4}, \city{M{\"u}nchen}, \postcode{80799}, \country{Germany}}}
\affil[3]{\orgdiv{QuantumDiamonds GmbH}, \orgaddress{\street{Friedenstr. 6}, \city{M{\"u}nchen}, \postcode{81671}, \country{Germany}}}
\affil[4]{\orgdiv{Schulich Faculty of Chemistry, Technion  - Israel Institute of Technology} \orgaddress{\street{}, \city{Haifa}, \postcode{32000}, \country{Israel}}}

\maketitle

\thispagestyle{empty}
\clearpage
\setcounter{page}{1}
\subsection*{Supplementary Note 1: Direct observation of spinlock phase accumulation dynamics}
As discussed in the main text, the phase $\theta_\mathrm{SL}$ accumulated during a spinlock pulse at the matching condition $\Omega_\mathrm{SL}=2\pi \nu_\mathrm{RF}$ is given by $\theta_\mathrm{SL}=\frac{1}{2}\gamma \hat{B}_\mathrm{RF} t_s$ with the gyromagnetic ratio $\gamma$, the magnetic RF field amplitude $\hat{B}_\mathrm{RF}$ and the spinlock duration $t_s$ \cite{rizzato_extending_2023}. In contrast to PDD sequences, where phase is accumulated on the xy-plane of the Bloch sphere in the first rotating frame, the spin vector moves in a spiral motion from the $(-y)$ towards the $(+y)$ axis during the spinlock pulse, as indicated by the black arrow in Supplementary Figure \ref{fig:SL_phase-accum} (a), and then back to the $(-y)$ axis (blue arrow).\\
Experimentally, the spin dynamics can be probed by leaving out the last $\pi/2$ projection pulse of the typical CDD sequence shown in the inset of Figure 2 (a) of the main text. In Supplementary Figure \ref{fig:SL_phase-accum} (b), the experimental result for a spinlock amplitude of $\Omega_\mathrm{SL}=2\pi \cdot 5 \, \mathrm{MHz}$ is shown, exhibiting fast oscillations at the Rabi frequency ($5 \, \mathrm{MHz}$) for a full evolution from the $(-y)$ axis towards the $(+y)$ axis (black) and back (blue). 
\begin{figure*}[ht]
  \centering
  {\includegraphics[width=0.8\textwidth]{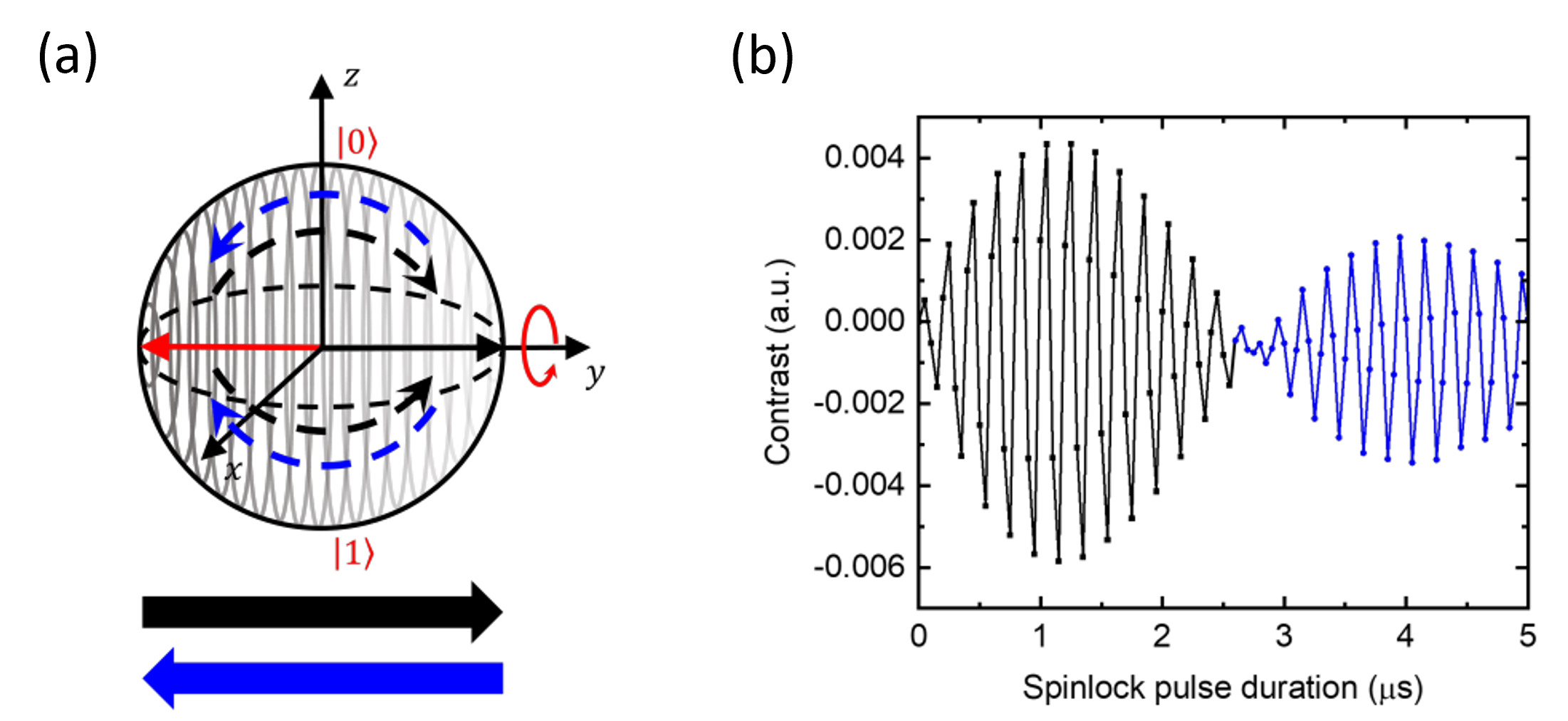}}
  \caption{\textbf{CDD spin dynamics.} (a) Schematic depiction of the spiraling motion in the first rotating frame during spinlock phase accumulation. (b) Direct observation of the spiraling dynamics by omitting the last $\pi/2$ projection pulse.}
 \label{fig:SL_phase-accum}
\end{figure*}
\clearpage

\subsection*{Supplementary Note 2: Microwave resonator design} 
\begin{figure*}[ht]
  \centering
  {\includegraphics[width=0.9\textwidth]{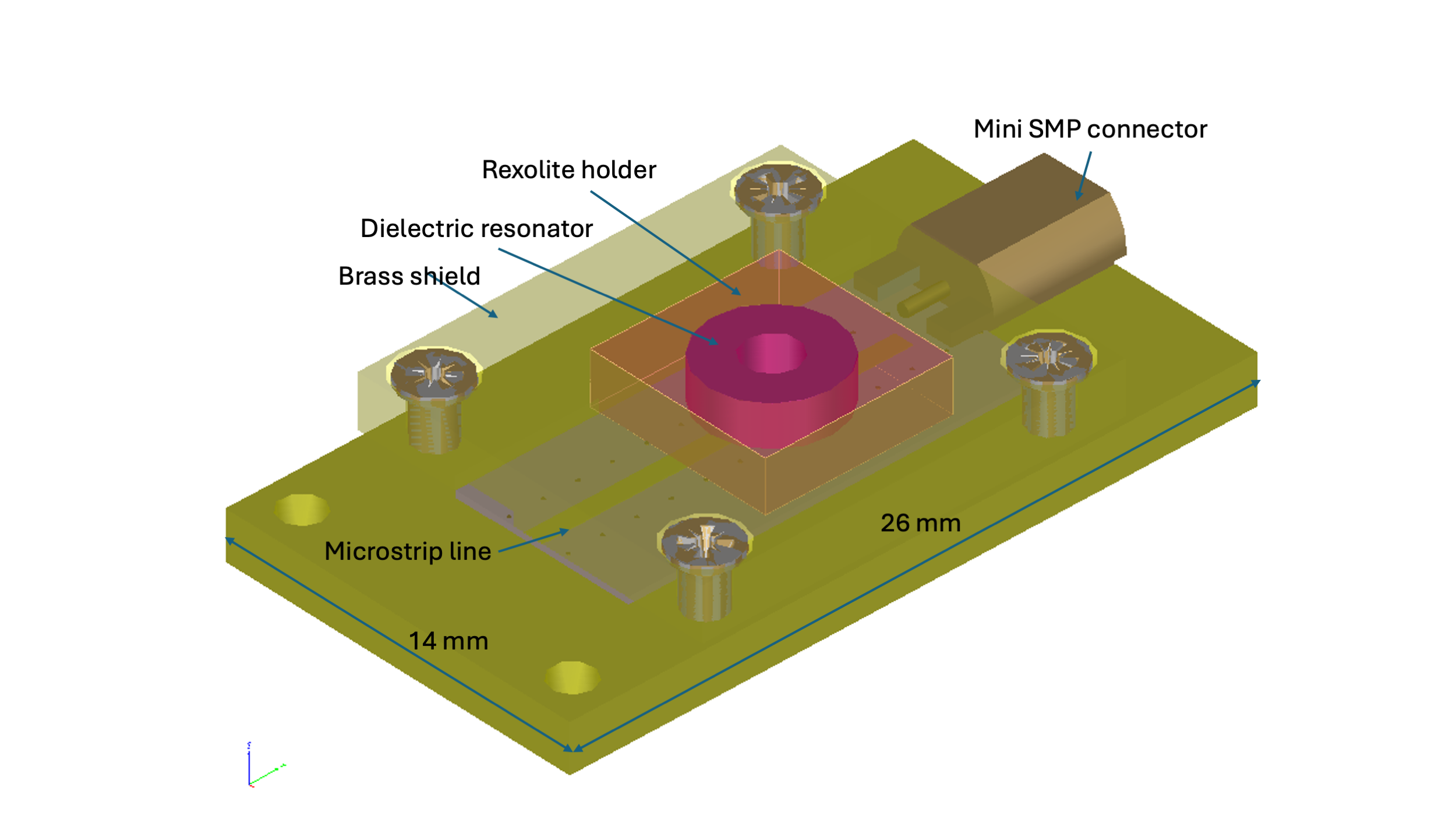}}
 \label{fig:SI_Resonator}
\end{figure*}

\begin{figure*}[ht]
  \centering
  {\includegraphics[width=0.9\textwidth]{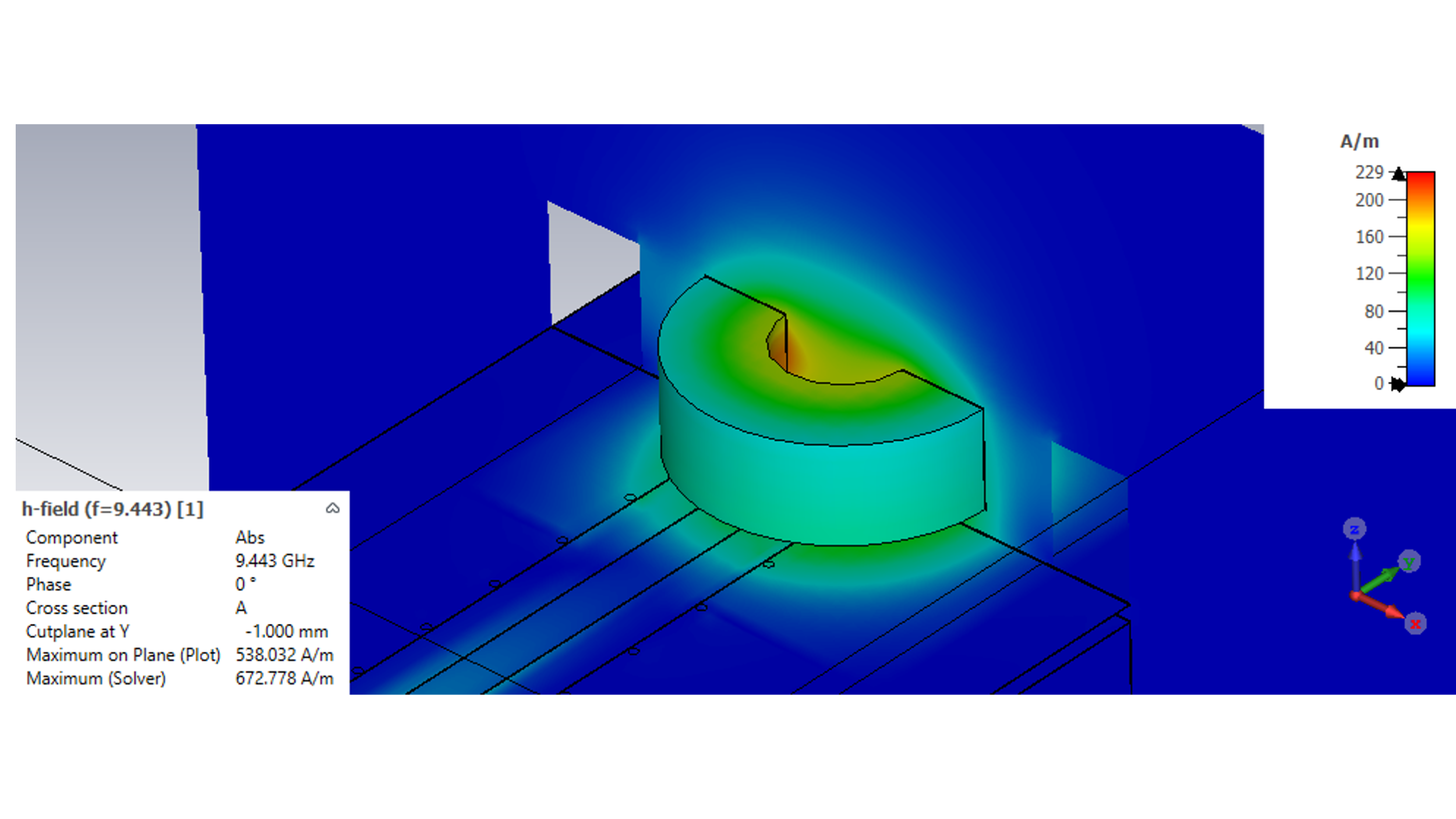}}
  \caption{\textbf{Resonator design.}}
 \label{fig:SI_Resonator2}
\end{figure*}

\subsection*{Supplementary Note 3: Rabi experiment}
\begin{figure*}[ht]
  \centering
  {\includegraphics[width=1\textwidth]{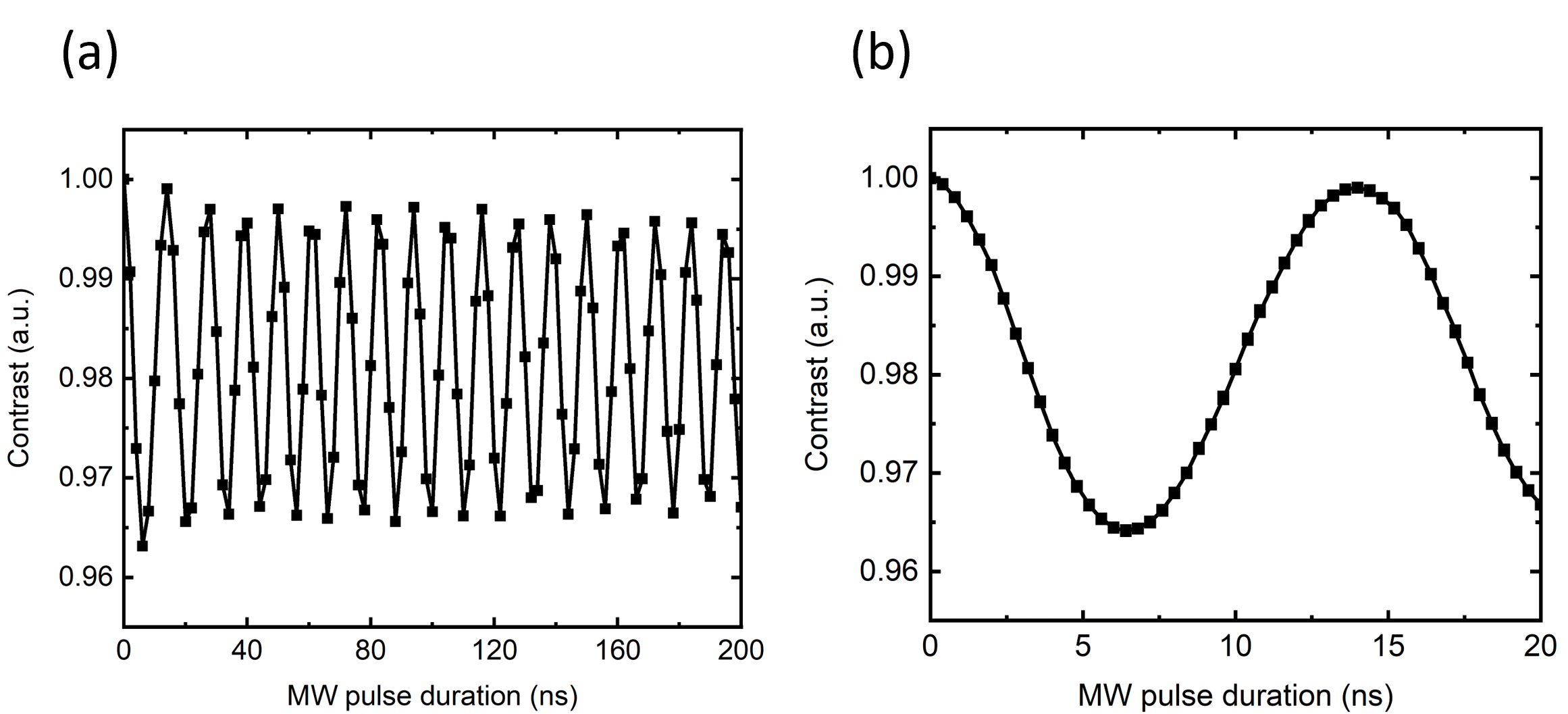}}
  \caption{\textbf{Rabi experiment.} Measured Rabi oscillations up to a MW pulse duration of $200 \, \mathrm{ns}$ (a) and $20 \, \mathrm{ns}$ (b).}
 \label{fig:rabi}
\end{figure*}

To determine the $\pi$ and $\pi/2$ pulse durations, a Rabi experiment was performed whereby the MW pulse duration was swept at the maximum MW power. Supplementary Figure \ref{fig:rabi} displays the measured Rabi-oscillations up to a pulse duration of $200 \, \mathrm{ns}$ (\ref{fig:rabi} (a)) and up to $20 \, \mathrm{ns}$ (\ref{fig:rabi} (b)). From this experiment, $t_{\pi/2}\approx3 \, \mathrm{ns}$ and $t_{\pi}\approx 6 \, \mathrm{ns}$ were determined, corresponding to a Rabi frequency of about $80 \, \mathrm{MHz}$ with a maximum contrast of approximately $4 \%$. Furthermore, Supplementary Figure \ref{fig:rabi} (a) exhibits almost no decay in contrast over a timeframe of $200 \, \mathrm{ns}$, indicating a homogeneous MW field over the laser spot (estimated spot diameter of $\sim 50 \, \upmu \mathrm{m}$) provided by the MW resonator.

\subsection*{Supplementary Note 4: Resonator ringdown time} 
\begin{figure*}[ht]
  \centering
  {\includegraphics[width=1\textwidth]{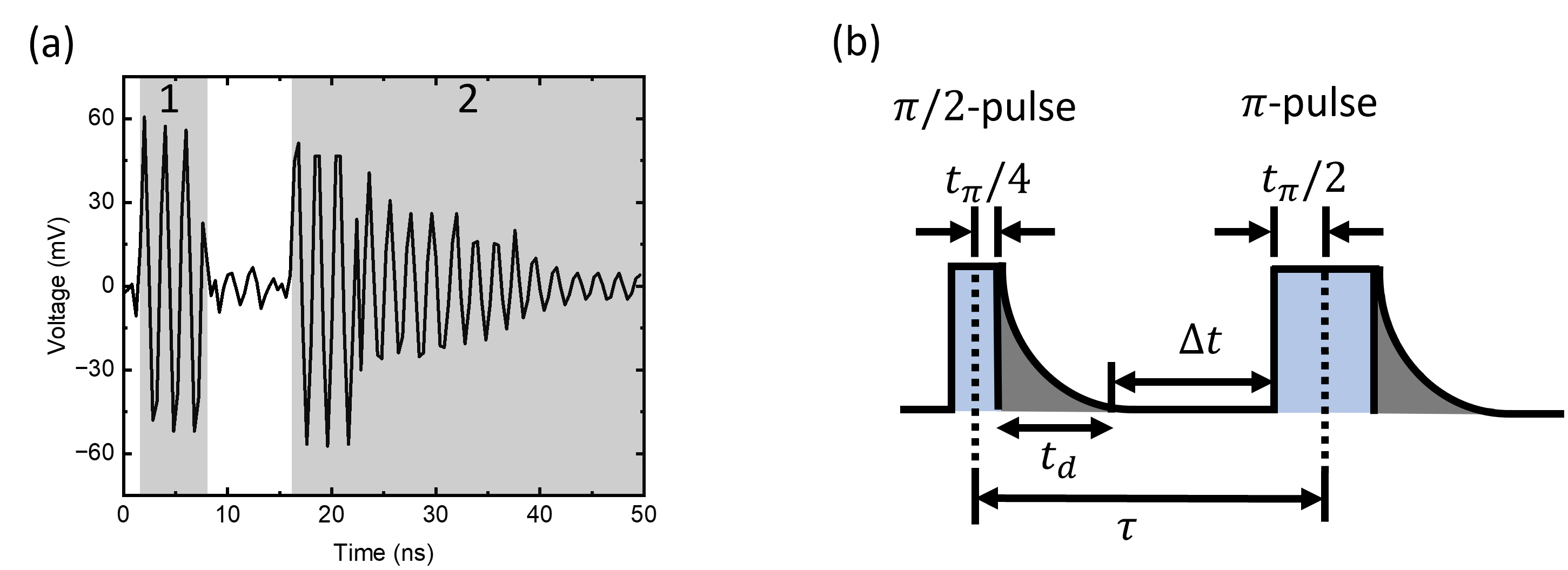}}
  \caption{\textbf{Microwave pulse characterization.} (a) The incoming $\pi$-pulse (marked area 1) is reflected behind the resonator (area 2) which exhibits a significantly long characteristic decay time. (b) Illustration of the relationship between pulse spacing $\tau$, $\pi$-pulse duration $t_\pi$, decay time $t_d$ and pulse separation $\Delta t$.}
 \label{fig:SI_ringdown}
\end{figure*}

The microwave pulses used to drive the NV spin states are not perfectly rectangularly shaped, but they exhibit a significant ringdown decay time due to the microwave resonator's Q-factor of approx. 720, which was measured with a vector network analyzer. The Q-factor together with the resonance frequency $f_0$ at $9.4 \, \mathrm{GHz}$ correspond to an estimated decay time $t_d$ of $\approx 25 \, \mathrm{ns}$ according to \cite{jackson1998, Heintze.2014}
\begin{equation}
    t_d=\frac{Q}{\pi f_0} \, .
\end{equation}
This has been verified experimentally by recording the (downmixed) reflected signal behind the resonator with an oscilloscope. Supplementary Figure \ref{fig:SI_ringdown} (a) shows an incoming $\pi$ pulse (marked area 1) with a set pulse duration of $\approx 6 \, \mathrm{ns}$ and its reflection behind the resonator (area 2). In contrast to the incoming pulse, the reflected signal displays a considerably long ringdown time which is in good agreement with the aforementioned calculated value.\\
As a consequence, an overlap of pulses for sensing frequencies at and above $10 \, \mathrm{MHz}$ occurs: As shown in Supplementary Figure \ref{fig:SI_ringdown} (b), the spacing $\tau$ between the first $\pi/2$ and $\pi$ pulse (measured from the center of the pulses) can be expressed as follows:
\begin{equation}
    \tau=\frac{3}{4} t_\pi +t_d+\Delta t \,
\end{equation}
where $t_\pi$ is the $\pi$-pulse duration and $\Delta t$ the separation between the pulses taking into account the decay time $t_d$. For $\Delta t<0$, the tail of the $\pi/2$-pulse overlaps with the subsequent $\pi$-pulse, negatively affecting the sensor's capability to manipulate the spin qubit accurately. The minimum separation without overlap is therefore $\Delta t=0$, corresponding to a spacing of $\tau_\mathrm{overlap}=\frac{3}{4} t_\pi+t_d\approx30 \, \mathrm{ns}$ for $t_\pi=6 \, \mathrm{ns}$ and $t_d=25 \, \mathrm{ns}$. The minimal sensing frequency at which the pulses overlap is given by the matching condition $\nu_\mathrm{overlap}=1/(4\tau_\mathrm{overlap}) \approx 8 \, \mathrm{MHz}$ which matches the observations in the main experiments (see Results in the main text).
\subsection*{Supplementary Note 5: Evidence for MW resonator heating effects} 
\begin{figure*}[ht]
  \centering
  {\includegraphics[width=0.5\textwidth]{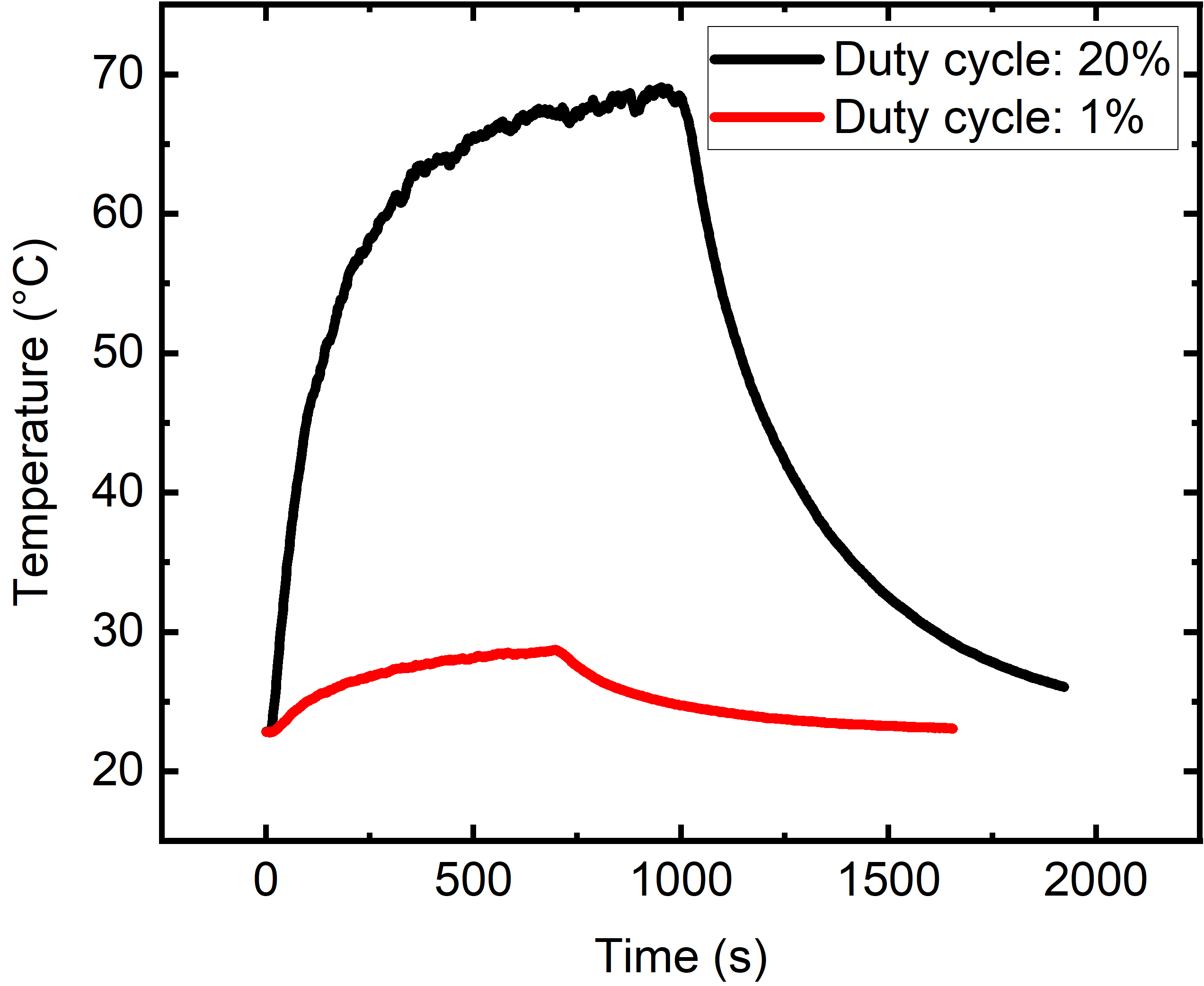}}
  \caption{\textbf{Resonator temperature for two different duty cycles.} The black curve corresponds to a set duty cycle of $20\%$ whereas the red curve results from a duty cycle of $1\%$.}
 \label{fig:res_heating}
\end{figure*}

The microwave resonator, connected to a $\sim 60 \, \mathrm{W}$ amplifier, generates strong and homogeneous MW fields required for high-frequency NV sensing. To investigate possible heating effects due to the strong MW power used, the temperature on the resonator next to the diamond position was probed using a temperature sensor (Thorlabs TSP01). The chosen MW power was at $60 \%$ of its maximum value and the tested duty cycles were $20\%$ ($100 \, \upmu \mathrm{s}$ MW duration and $400 \, \upmu \mathrm{s}$ dead time) and $1\%$ ($5 \, \upmu \mathrm{s}$ MW duration and $400 \, \upmu \mathrm{s}$ dead time). The duty cycle in the experiments described in the main text is $\approx 0.5 \%$ ($5 \, \upmu \mathrm{s}$ MW duration and $1.1 \, \mathrm{ms}$ dead time).\\
As evident from Supplementary Figure \ref{fig:res_heating}, the recorded temperature rises from just above $20^\circ \mathrm{C}$ to $\approx 70^\circ \mathrm{C}$ in case of a duty cycle of $20\%$ and to $\approx 30 ^\circ \mathrm{C}$ for the $1\%$ duty cycle, indicating significant heating effects which might lead to instabilities that may result in line broadening.\\
The results displayed in Supplementary Figure 2 of the main text have been reproduced with a longer spinlock pulse duration of $t_s=20 \, \upmu \mathrm{s}$ and dead time of $1.5 \, \mathrm{ms}$ corresponding to a duty cycle of $\approx 1\%$. The applied RF field amplitude (see Supplementary Note 4) has been scaled such that the accumulated phase during the spinlock duration is the same for both experiments ($t_s=5\, \upmu\mathrm{s}$ and $t_s=20\, \upmu \mathrm{s}$). The comparison in Supplementary Figure \ref{fig:5us-vs-20us}, exhibits a considerably lower contrast and lower achievable maximum RF sensing frequency for a higher duty cycle. As already mentioned, suspected heating effects may deteriorate the CDD sensing capabilities.
\begin{figure*}[ht]
  \centering
  {\includegraphics[width=0.9\textwidth]{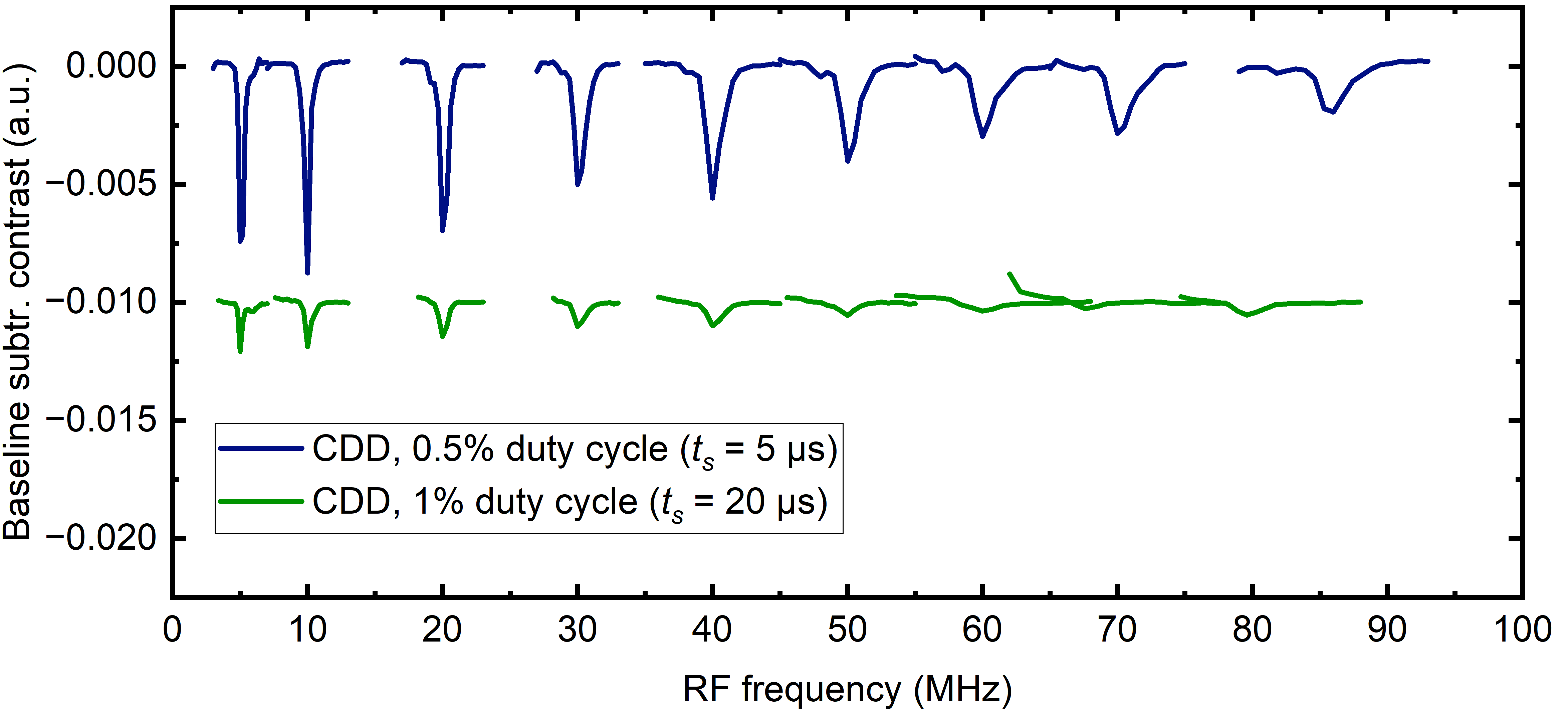}}
  \caption{\textbf{Frequency response of the CDD detection scheme with two differing duty cycles.} The blue curve (from Figure 2 in the main text) corresponds to an active sensor time $t_s$ of $5 \, \upmu \mathrm{s}$ and a dead time of $1.1 \, \mathrm{ms}$ (duty cycle of $\approx 0.5 \%$) and the green curve to a spinlock duration of $20 \, \upmu \mathrm{s}$ a dead time of $1.5 \, \mathrm{ms}$ (duty cycle of $\approx 1 \%$).}
 \label{fig:5us-vs-20us}
\end{figure*}

\subsection*{Supplementary Note 6: Calibration of the RF amplitude $\hat{B}_\mathrm{RF}$} 
\begin{figure*}[ht]
  \centering
  {\includegraphics[width=0.5\textwidth]{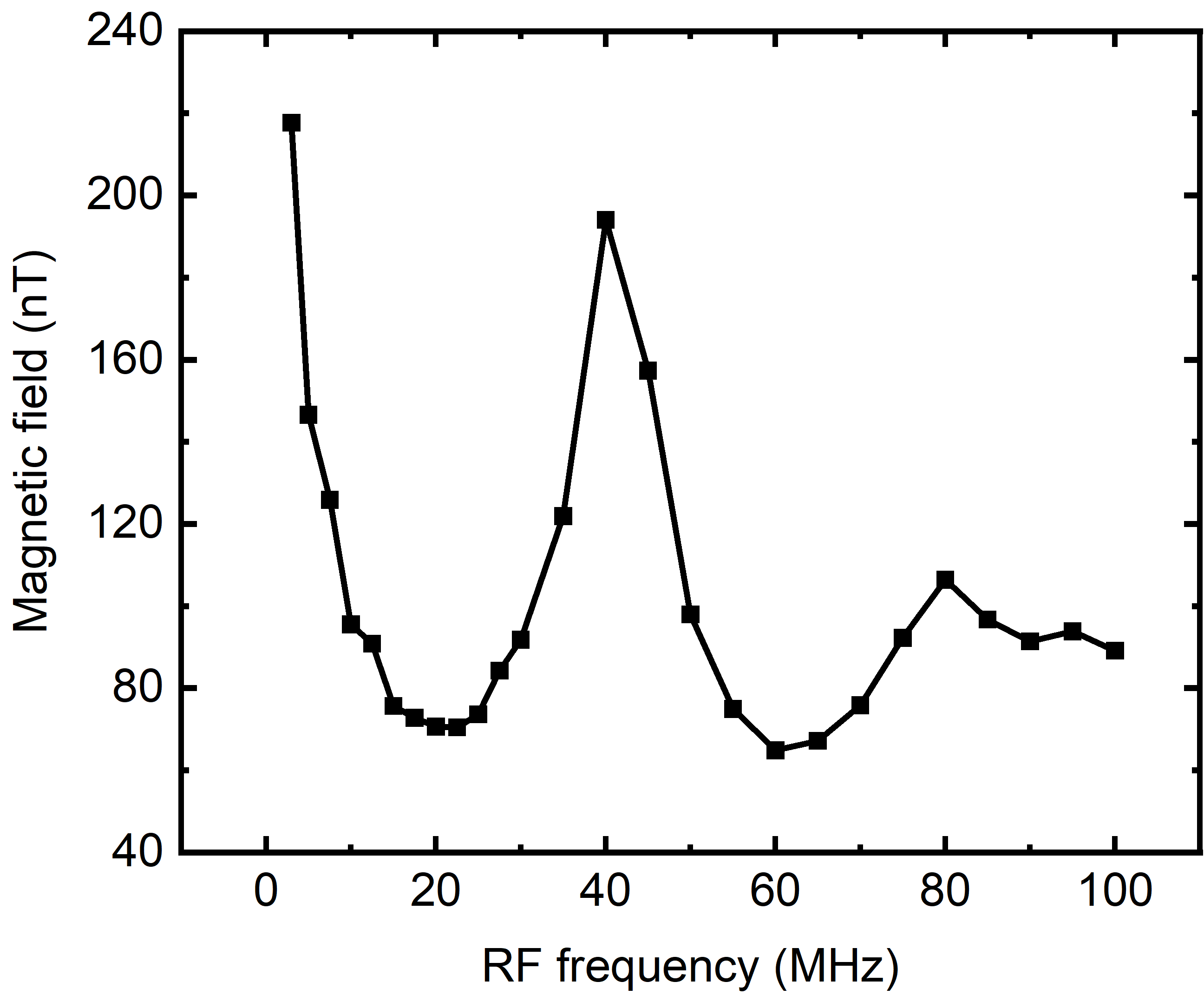}}
  \caption{\textbf{Calibration of the RF amplitude $\hat{B}_\mathrm{RF}$.} Due to the frequency dependency of the RF coil's impedance, the amplitude of the RF source was adjusted for a constant magnetic field output for each sensing frequency in the main experiments.}
 \label{fig:coil_calib}
\end{figure*}

The RF signal sensed by the NV detector is provided by a self-assembled coil near the diamond position. The impedance of the RF coil is frequency dependent, thus, the output voltage of the RF signal generator was calibrated such that the sensed RF amplitude $\hat{B}_\mathrm{RF}$ is approximately the same for each measurement frequency. This was achieved by placing a pick-up coil (Beehive Electronics 100A) near the RF coil and by reading out the pick-up coil's induced voltage amplitude via an oscilloscope at a constant RF signal source voltage for each applied frequency. Supplementary Figure \ref{fig:coil_calib} shows the magnetic field output of the RF coil in dependence of the RF frequency at the position of the pick-up coil. From this, the voltage of the RF source was adjusted such that the amplitude $\hat{B}_\mathrm{RF}$ is constant over all applied RF frequencies.

\clearpage
\bibliography{bibliography.bib}